\documentclass{article}
\usepackage{fullpage}

\usepackage{microtype}
\usepackage{graphicx}
\usepackage{subcaption}
\usepackage{booktabs} %

\usepackage{hyperref}

\usepackage{amsmath}
\usepackage{amssymb}
\usepackage{mathtools}
\usepackage{amsthm}

\usepackage{array}
\usepackage{textcomp}
\usepackage{stfloats}
\usepackage{url}
\usepackage{verbatim}
\usepackage{graphicx}

\usepackage{bm}
\usepackage{booktabs}
\usepackage{siunitx}  %
\usepackage{multirow} %
\usepackage[flushleft]{threeparttable}
\usepackage{enumitem}
\usepackage{xcolor}
\usepackage{colortbl}
\usepackage{mathtools}
\usepackage{pifont}
\usepackage{amsthm}
\usepackage{arydshln}		%
\usepackage{makecell}
\usepackage{fontawesome5}

\def\min{\mathop{\mathsf{min}}}

\def\Abm{{\bm{A}}}

\def\Fbm{{\bm{F}}}

\def\Sbm{{\bm{S}}}

\def\Abm{{\bm{A}}}

\def\Fbm{{\bm{F}}}

\def\Tsf{{\mathsf{T}}}

\def\Tsf{{\mathsf{T}}}

\def\lim{\mathop{\mathrm{lim}}} %
\def\min{\mathop{\mathrm{min}}}

\def\xbm{{\bm{x}}}

\def\ybm{{\bm{y}}}

\def\Abm{{\bm{A}}}

\def\Fbm{{\bm{F}}}

\def\Sbm{{\bm{S}}}

\def\Mbm{{\bm{M}}}

\def\Tsf{{\mathsf{T}}}

\definecolor{lightgreen}{rgb}{.9,1,.9}
\definecolor{lightred}{rgb}{1,.415,.415}
\definecolor{lightblue}{rgb}{.415,.415,1}

\newcolumntype{L}[1]{>{\raggedright\arraybackslash}p{#1}}
\newcolumntype{C}[1]{>{\centering\arraybackslash}p{#1}}
\newcolumntype{R}[1]{>{\raggedleft\arraybackslash}p{#1}}

\usepackage[capitalize,noabbrev]{cleveref}

\theoremstyle{plain}

\theoremstyle{definition}

\theoremstyle{remark}

\usepackage[textsize=tiny]{todonotes}

\begin{document}

\title{A Unified Framework for Multimodal Image Reconstruction and Synthesis using Denoising Diffusion Models}
\date{}

\author{Weijie Gan$^{*,1}$, Xucheng Wang$^{*,1}$, Tongyao Wang$^2$, Wenshang Wang$^2$, Chunwei Ying$^2$,\\ Yuyang Hu$^3$, Yasheng Chen$^4$, Hongyu An$^{2,3,4,5,6}$ and Ulugbek S. Kamilov$^7$\\ \\
\small $^1$Department of Computer Science and Engineering, Washington University in St. Louis, St. Louis, MO, USA\\
\small $^2$Mallinckrodt Institute of Radiology, Washington University in St. Louis, St. Louis, MO USA\\
\small $^3$Department of Electrical and Systems Engineering, Washington University in St. Louis, St. Louis, MO, USA\\
\small $^4$Department of Neurology, Washington University in St. Louis, St. Louis, MO USA\\
\small $^5$Department of Biomedical Engineering, Washington University in St. Louis, St. Louis, MO USA\\
\small $^6$Division of Biology and Biomedical Sciences, Washington University in St. Louis, St. Louis, MO, USA\\
\small $^7$Department of Electrical and Computer Engineering, University of Wisconsin–Madison, Madison WI, USA\\
$^{*}$\small These authors contributed equally.\\
\small\texttt{$\{$weijie.gan, david.w, tongyaow, wenshangwang$\}$@wustl.edu}\\ 
\small\texttt{$\{$chunwei.ying, h.yuyang, yasheng.chen, hongyuan$\}$@wustl.edu}\\ 
\small\texttt{$\{$kamilov$\}$@wisc.edu}
}

\maketitle

\begin{abstract}
  Image reconstruction and image synthesis are important for handling incomplete multimodal imaging data, but existing methods require various task-specific models, complicating training and deployment workflows. We introduce Any2all, a unified framework that addresses this limitation by formulating these disparate tasks as a single virtual inpainting problem. We train a single, unconditional diffusion model on the complete multimodal data stack. This model is then adapted at inference time to ``inpaint'' \textit{all} target modalities from \textit{any} combination of inputs of available clean images or noisy measurements. We validated Any2all on a PET/MR/CT brain dataset. Our results show that Any2all can achieve excellent performance on both multimodal reconstruction and synthesis tasks, consistently yielding images with competitive distortion-based performance and  superior perceptual quality over specialized methods. 
\end{abstract}

\section{Introduction}

Multimodal imaging (MMI), the practice of integrating data from multiple distinct sensors, is a powerful paradigm for a comprehensive understanding across numerous scientific and engineering domains~\cite{zhan2023multimodal,marti2010multimodality}. For instance, in autonomous driving systems, fusing information from cameras, LiDAR, and radar is critical for robust perception and navigation~\cite{xiao2020multimodal}. This paper focuses on the application of this paradigm within medical imaging, where MMI has become indispensable. By integrating modalities like magnetic resonance imaging (MRI), computed tomography (CT), and positron emission tomography (PET), clinicians can gather diverse anatomical, functional, and pathological information from a single patient examination. For example, combining T1- and T2-weighted MRI scans offers a more complete picture by providing both anatomical and pathological details simultaneously~\cite{any2all_Havaei2017}.

Despite its diagnostic power, the practical application of MMI is often hindered by significant challenges, including high financial costs, extended scan times, and concerns about patient exposure to ionizing radiation. To mitigate these challenges, two common strategies are employed in practice: \emph{(a)} accelerating the acquisition process for certain modalities, or \emph{(b)} skipping the acquisition of some modalities altogether. These strategies result in incomplete data, leading to either corrupted images from undersampled scans or entirely missing images. To overcome this data gap, computational techniques, especially those based on machine learning (ML), have been widely adopted. Image reconstruction algorithms are used to enhance image quality from the corrupted, accelerated scans~\cite{Wang.etal2020,Ongie.etal2020,mccann2017convolutional}, while image synthesis (or image translation) methods are used to generate the missing modalities from available data~\cite{Dayarathna.etal2024,Zhan.etal2023}.

However, existing ML approaches predominantly rely on developing and training distinct, specialized models for each specific task. This task-specific methodology complicates both the training and deployment pipelines and lacks the flexibility to adapt to diverse clinical scenarios at inference time. To address this issue, in this paper, we propose a unified generative framework, termed \emph{Any2all}, that allows mapping \emph{any} combination of acquired data to \emph{all} desired multimodal images. Our key contributions are threefold:

\begin{enumerate}[itemsep=1pt, parsep=1pt, topsep=1pt]
\item We introduce a novel formulation that unifies multimodal image reconstruction and synthesis. We view these tasks as a \emph{virtual inpainting problem}, where the goal is to fill in missing or corrupted information within a complete images stack of target modalities. This perspective allows us to train a single, unconditional diffusion model on the complete images stack of target modalities and use it in any specific downstream task.

\item We propose two novel, task-adaptive sampling algorithms to solve this virtual inpainting problem at inference time: \emph{multimodal posterior sampling (MPS)} and \emph{multimodal decomposition sampling (MDS)}. Both algorithms iteratively guide the generation process by applying a correction that enforces consistency between the model's denoised predictions and the available data, whether clean images or noisy measurements. Their key difference lies in the correction strategy: MPS steers the sampling process with a gradient-based correction, while MDS directly refines the predicted clean image at each step. Together, these complementary samplers allow our single pre-trained model to be flexibly adapted to any reconstruction or synthesis task.

\item We perform a comprehensive validation of the Any2all pipeline on a dataset with MRI, CT, and PET modalities. Through extensive experiments, we demonstrate that our unified framework achieves excellent results in both image reconstruction and synthesis tasks. It consistently yields superior perceptual quality and competitive distortion-based performance when compared to various specialized baselines.
\end{enumerate}

\begin{figure*}
  \centering
  \includegraphics[width=\textwidth]{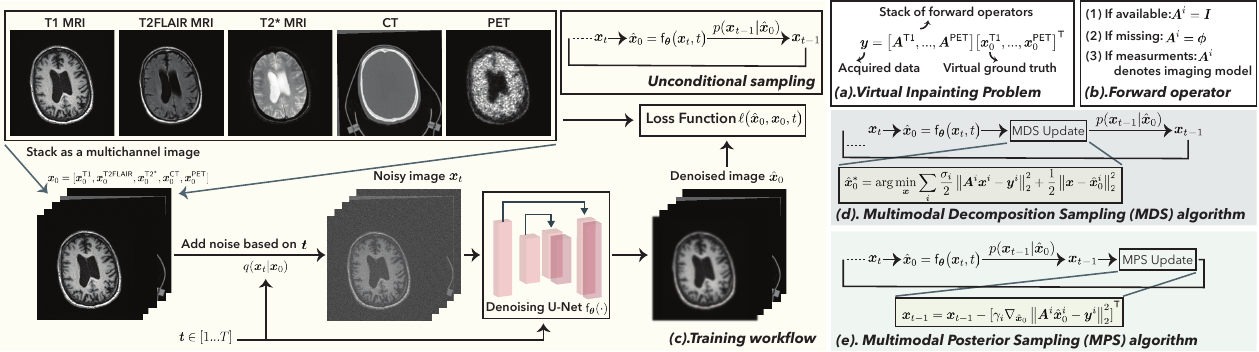}
  \caption{An illustration of the training and inference pipelines for Any2all. \textit{\textbf{(a, b)}}: Our framework is founded on the concept of a \emph{unified virtual inpainting problem}, which treats diverse image reconstruction and synthesis tasks as the common goal of restoring missing or corrupted information within a complete set of multimodal images. \textit{\textbf{(c)}}: To solve this, we train a single, unconditional diffusion model that serves as a powerful generative prior. {\textbf{(d, e)}}: During inference, we propose task-adaptive sampling algorithms, namely \emph{MPS} and \emph{MDS}, to steer the generative process by enforcing constraints from the available data. This figure shows the ability of Any2all to use a single pre-trained model to map \emph{any} input data to \emph{all} clean, multimodal images.}
  \label{fig:any2all-method}
\end{figure*}

\section{Related Work}

MMI relies on two fundamental but distinct computational tasks to generate and supplement missing data: image reconstruction and image synthesis. While both leverage computational methods to produce images, they address different problems in the workflow.

\subsection{Multimodal Image Reconstruction}

Image reconstruction is the process of generating a human-interpretable image from raw sensor data, which is fundamentally an ill-posed inverse problem. The goal is to produce high-quality images for clinical use while minimizing factors like radiation dose and scan time, which often requires reconstruction from undersampled or noisy data. Historically, single-contrast reconstruction algorithms evolved through three main paradigms~\cite{Wang.etal2020}: analytic methods based on mathematical inversions, iterative methods that use a numerical forward model and feedback loop to progressively refine an image~\cite{Hu.etal2012,Elad.Aharon2006}, and deep learning (DL) methods that learn the inversion from data. These DL methods, which range from deep convolutional neural networks~\cite{Chen.etal2017,McCann.etal2017,Gan.etal2022} to model-based deep learning methods~\cite{Monga.etal2021,kamilov2023plug,gan2023block}, have become the mainstream solution.

Building on this foundation, multimodal image reconstruction leverages one or more high-quality reference images (\emph{e.g.,} T1-weighted MRI) to provide valuable supplementary information for reconstructing a target image (\emph{e.g.,} a T2-weighted MRI). A primary challenge is to effectively fuse this information without introducing interference from inconsistent features between modalities. Prevailing deep learning strategies include decomposition-based approaches and generative prior approaches.
Decomposition-based methods are motivated by the observation that different contrasts share common structures but have unique intensity profiles. These methods explicitly decompose a reference image into a ``common'' and a ``unique'' component~\cite{Lei.etal2023,Zhou.Zhou2020,Xiang.etal2019,Feng.etal2021}. Only the shared, common information is then transferred to guide the target reconstruction, leading to an interpretable, model-driven deep network.
Generative prior approaches leverages diffusion models to act as a powerful generative prior, producing high-frequency details or other latent features to guide the reconstruction~\cite{Mao.etal2023a,Levac.etal2023a,Li.etal2024a}. In this multimodal setup, the reconstruction network is guided by two streams of information: \emph{(a)} complementary features extracted and fused from the available reference and target modalities, and \emph{(b)} the learned generative prior from the diffusion model. Fusing these two sources allows the network to restore the target image with high-frequency details that are consistent with both the reference anatomy and the learned data distribution. While standard diffusion models can be computationally intensive, recent works improve efficiency by applying the diffusion process within a highly compact latent space~\cite{Li.etal2024a}, allowing for the generation of accurate priors in very few iterations.

\subsection{Multimodal Image Synthesis}

Image synthesis, or image-to-image translation, generates a new image of a target modality based on one or more available source images~\cite{Dayarathna.etal2024,Zhan.etal2023}. This field is dominated by deep generative models that have proven highly effective at learning the complex, non-linear mappings between modalities. The core generative architectures include generative adversarial networks (GANs) and diffusion models.
GANs utilize a two-player system where a generator creates synthetic images and a discriminator attempts to distinguish them from real images~\cite{Zhou.etal2023,Zimmermann.etal2022}. Through this adversarial competition, the generator learns to produce high-fidelity images.
Diffusion models operate by reversing a diffusion process~\cite{Ozbey.etal2023,Wang.etal2024a}. A forward process gradually adds noise to an image until it becomes unrecognizable, and a learned reverse process then progressively denoises a random input to generate a clean, new image. Diffusion models are known for generating high-quality and diverse samples.

For multimodal image synthesis, these foundational generative models are used in a conditional framework where the available source modalities guide the generation of the missing target modalities. The strategies for this conditional synthesis can be categorized by their mapping schemes. Early works focused on ``one-to-one''~\cite{Pan.etal2024a} or ``many-to-one''~\cite{Liu.etal2023c,Jiang.etal2023} translation, where models were designed for specific, predefined source-target pairs. To handle more flexible clinical scenarios, ``many-to-many''~\cite{Chartsias.etal2018,Meng.etal2024a} frameworks were developed to generate any combination of missing modalities simultaneously. These methods often work by pre-imputing missing modalities with zeros or fusing features from available modalities using a predefined function.
To improve synthesis quality, recent works have focused on advanced network architectures. ResViT~\cite{Dalmaz.etal2022}, for example, introduced a hybrid model that combines the local precision of CNNs with the ability of Vision Transformers to capture long-range, contextual relationships in medical images. The diffusion-based model M2DN~\cite{Meng.etal2024a} reframes the problem from ``cross-modal translation'' to ``progressive whole-modality inpainting'', where missing modalities are treated as noise and synthesized jointly with the self-reconstruction of available ones to better construct a common latent space.

\section{Method}

\subsection{Problem Formulation}
\label{subsec:any2all-problem-formulation}
To address the challenges of incomplete multimodal data, we propose a unified generative framework, which we term \emph{Any2all}. The name reflects the core capability of our method: to take \emph{any} combination of available data as input and generate \emph{all} desired modalities as output. This is achieved by tackling the problem from a novel viewpoint that unifies both multimodal image reconstruction and image synthesis into a single \emph{virtual inpainting problem}.

First, we define a ``virtual'' ground truth, $\bm{x}$, as a stack of all $n$ desired high-quality modality images:
\begin{equation}
    \bm{x} = [\bm{x}_1, \bm{x}_2, \dots, \bm{x}_n]^\Tsf
\end{equation}
For instance, in a clinical brain scan scenario involving multiple MRI contrasts as well as CT and PET, this stack could be represented as $\bm{x} = [\bm{x}^{\text{T1}}, \bm{x}^{\text{T2FLAIR}}, \bm{x}^{\text{T2*}}, \bm{x}^{\text{CT}}, \bm{x}^{\text{PET}}]^T$.

Under this formulation, any combination of available data at testing time can be considered a linear measurement of this complete virtual ground truth. We model this relationship as:
\begin{equation}
\bm{y} = \bm{A}\bm{x}
\end{equation}
where $\bm{y}$ represents the collection of all acquired data and $\bm{A}$ is a diagonal matrix of measurement operators, $\bm{A} = \text{diag}(\bm{A}_1, \dots, \bm{A}_n)$. The specific form of each operator $\bm{A}_i$ depends on the status of the $i$-th modality:
    (1) The $i$-th modality is \emph{available and clean:} The operator is the identity matrix, $\bm{A}_i = \bm{I}$. The measurement $\bm{y}_i$ is simply the ground truth image $\bm{x}_i$, 
    (2) The $i$-th modality is \emph{missing:} The operator is zero, $\bm{A}_i = 0$. This indicates that no information is available for this modality.
    and (3) The $i$-th modality has \emph{noisy or undersampled measurements:} The operator $\bm{A}_i$ represents the forward model that maps the clean image to the acquired measurements (e.g., an undersampling mask in MRI k-space).

This Any2all formulation is powerful because it casts diverse and seemingly distinct tasks into a single, consistent inverse problem. The flexibility of the measurement operator $\bm{A}$ allows for any combination of inputs, and the objective is always to recover the full ground truth stack $\bm{x}$. This goal of mapping from an arbitrary set of inputs to the complete set of all modalities is, fundamentally, an \emph{inpainting problem}.

\subsection{Model Training}

Based on our formulation of multimodal imaging as a virtual inpainting problem, the Any2all framework requires only a single generative model to be trained. We employ an unconditional denoising diffusion probabilistic model (DDPM)~\cite{Ho.etal2020} that learns the underlying data distribution of the complete ``virtual'' ground truth stack, $\bm{x}_0$. This training process is entirely task-agnostic, meaning the model learns a general prior over the complete multimodal data without knowledge of any specific reconstruction or synthesis task.

As illustrated in Figure~\ref{fig:any2all-method}, the training is based on the standard DDPM framework, which consists of a fixed forward diffusion process and a learned reverse denoising process.
The forward process is a fixed Markov chain that gradually adds Gaussian noise to the clean data stack $\bm{x}_0$ over a series of $T$ timesteps. The data at step $t$ is generated from the data at step $t-1$ according to a Gaussian transition:
$$
q(\bm{x}_t|\bm{x}_{t-1}) = \mathcal{N}(\bm{x}_t; \sqrt{1-\beta_t}\bm{x}_{t-1}, \beta_t\bm{I})
$$
where $\{\beta_t\}_{t=1}^T$ is a predefined variance schedule that controls the noise level at each step. A key property of this process is that we can sample a noisy version $\bm{x}_t$ at any arbitrary timestep $t$ directly from the original clean data $\bm{x}_0$ in a closed form:
$$
q(\bm{x}_t|\bm{x}_0) = \mathcal{N}(\bm{x}_t; \sqrt{\bar{\alpha}_t}\bm{x}_0, (1-\bar{\alpha}_t)\bm{I})
$$
where $\alpha_t := 1 - \beta_t$ and $\bar{\alpha}_t := \prod_{s=1}^t \alpha_s$.

The goal of training is to learn the reverse process, $p_{\bm{\theta}}(\bm{x}_{t-1}|\bm{x}_t)$, which can gradually denoise a sample starting from pure Gaussian noise $\bm{x}_T \sim \mathcal{N}(0, \bm{I})$ back to a clean data sample $\bm{x}_0$. This is achieved by training a neural network, $\bm{\epsilon}_{\bm{\theta}}$, to predict the noise component $\bm{\epsilon}$ that was added to the clean data stack at a given timestep $t$. The network takes the noisy data stack $\bm{x}_t$ and the timestep $t$ as input and is optimized via a simplified mean-squared error objective function between the true and predicted noise:
$$
\mathcal{L} = \mathbb{E}_{t, \bm{x}_0} \left[ ||\bm{\epsilon} - \bm{\epsilon}_{\bm{\theta}}(\sqrt{\bar{\alpha}_t}\bm{x}_0 + \sqrt{1-\bar{\alpha}_t}\bm{\epsilon}, t)||_2^2 \right]
$$
where $t$ is sampled uniformly from $\{1, ..., T\}$, $\bm{x}_0$ is a clean data stack from the training set, and $\bm{\epsilon}$ is a random noise sample. By training the network $\bm{\epsilon}_{\bm{\theta}}$ on this objective, the model learns a generative prior for the entire multimodal data distribution.

\subsection{Model Inference}

With the unconditional diffusion model $ \bm{\epsilon}_\theta$ trained on the complete data stack $\bm{x}_0$, we perform inference by adapting the standard reverse sampling process to solve the inpainting problem defined in Section~\ref{subsec:any2all-problem-formulation}. The core idea is to guide the generation process at each timestep $t$ using the known measurements $\bm{y}$, ensuring the final output $\bm{x}_0$ is consistent with the available data. This guidance is achieved by introducing a correction step after the model predicts the denoised image stack. We adapt two powerful, state-of-the-art sampling strategies for this purpose, namely \emph{multimodal posterior sampling (MPS)} and \emph{multimodal decomposition sampling (MDS)}. By leveraging these two complementary sampling strategies, our single Any2all model can be flexibly adapted at inference time to solve a wide range of multimodal imaging tasks.

\subsubsection{Multimodal Posterior Sampling}
This first approach is inspired by diffusion posterior sampling (DPS)~\cite{Chung.etal2023}, a general framework for solving noisy inverse problems. MPS approximates posterior sampling by guiding the reverse diffusion process with a gradient that corrects for data inconsistency. 
The iterative formulation at each reverse step $t$ is: (1) \emph{predict denoised stack:} First, predict the clean "virtual" ground truth $\hat{\bm{x}}_0$ from the current noisy sample $\bm{x}_t$ using the trained network $\bm{\epsilon}_\theta$:
    $$
    \hat{\bm{x}}_0 = \frac{1}{\sqrt{\bar{\alpha}_t}}(\bm{x}_t - \sqrt{1-\bar{\alpha}_t}\bm{\epsilon}_\theta(\bm{x}_t, t))
    $$, (2) \emph{compute guidance gradient:} Calculate the gradient of the measurement error with respect to the current noisy sample $\bm{x}_t$:
    $$
    \bm{g}_t = \nabla_{\bm{x}_t} ||\bm{y} - \bm{A}\hat{\bm{x}}_0||_2^2
    $$, and (3) \emph{perform corrected update:} The next sample $\bm{x}_{t-1}$ is computed by taking a standard unconditional reverse diffusion step (e.g., a DDPM step) and then shifting it by the consistency gradient:
    $$
    \bm{x}_{t-1} = p_\theta(\bm{x}_{t-1}|\bm{x}_t) - \zeta_t \bm{g}_t\ ,
    $$
    where $p_\theta(\cdot|\cdot)$ is the unconditional reverse transition and $\zeta_t$ is a step size controlling the guidance strength.

\subsubsection{Multimodal Decomposition Sampling}
Our second approach is derived from decomposed diffusion sampling (DDS)~\cite{chung2024decomposed}, a technique that avoids expensive gradients by decomposing the sampling step and applying corrections in the clean image domain. 
The core insight is to decompose the DDIM sampling~\cite{Song.etal2021a} update into a denoised signal component and a noise component. The data consistency correction is applied only to the denoised signal. The iterative formulation at each step $t$ is: (1)
    \emph{predict denoised stack:} As before, predict the clean stack $\hat{\bm{x}}_0$ from the current noisy sample $\bm{x}_t$,
    (2) \emph{correct denoised stack:} The corrected estimate, $\hat{\bm{x}}^*_{0}$, is found by solving the following optimization problem in the clean image domain:
    $$
    \hat{\bm{x}}^*_{0} = \underset{\bm{x}}{\arg\min} \left( \frac{1}{2} ||\bm{y} - \bm{A}\bm{x}||_2^2 + \frac{\lambda}{2} ||\bm{x} - \hat{\bm{x}}_0||_2^2 \right)\ .
    $$
    This objective function balances data consistency with the measurements $\bm{y}$ (the first term) against fidelity to the diffusion model's prediction $\hat{\bm{x}}_0$ (the second term), with $\lambda$ as a regularization parameter. This standard quadratic optimization problem can be solved very efficiently using numerical methods such as the conjugate gradient (CG) algorithm. When a clean modality $\bm{x}_i$ is available, a special case of this correction is a direct replacement of the corresponding slice in $\hat{\bm{x}}_0$ (\emph{i.e.,} $\lambda_i \rightarrow 0$), and (3) \emph{construct next sample:} Recombine the corrected clean estimate $\hat{\bm{x}}^*_{0}$ with the appropriate noise component from the DDIM formulation to get the next sample $\bm{x}_{t-1}$:
    $$
    \bm{x}_{t-1} = \sqrt{\bar{\alpha}_{t-1}}\hat{\bm{x}}^*_{0} + \sqrt{1-\bar{\alpha}_{t-1}-\eta^2\tilde{\beta}_t^2} \cdot \bm{\epsilon}_\theta(\bm{x}_t,t) + \eta\tilde{\beta}_t\bm{\epsilon}\ ,
    $$
    where $\eta$ controls the level of stochasticity and $\bm{\epsilon} \sim \mathcal{N}(0, \bm{I})$.

\begin{figure*}
  \centering
  \includegraphics[width=\textwidth]{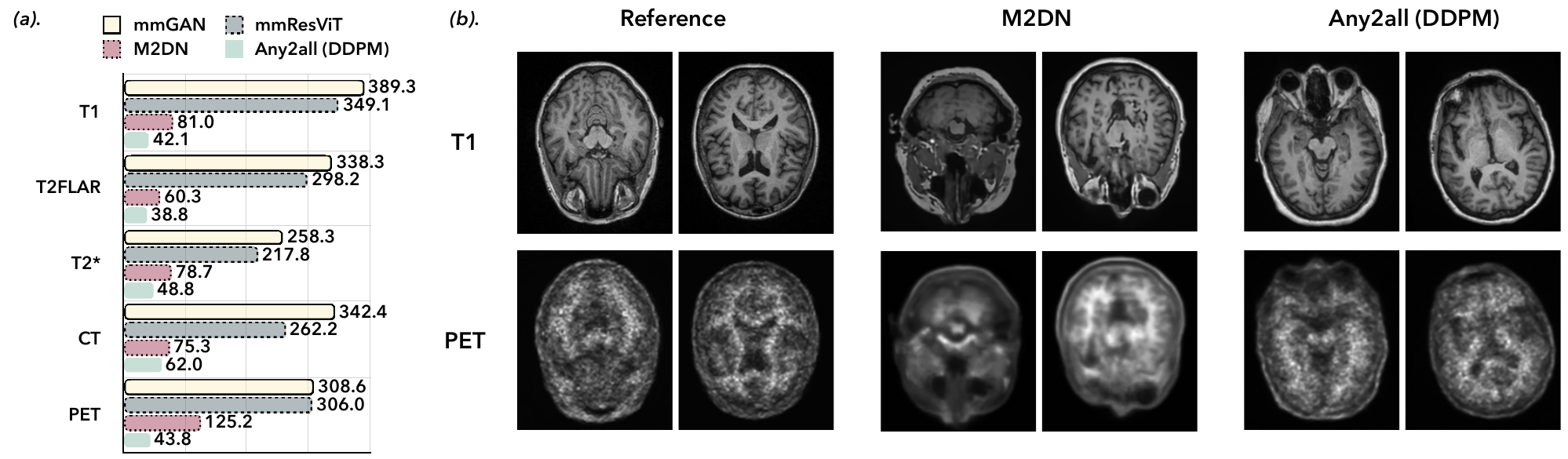}
  \caption{Qualitative and quantitative evaluation of unconditional image generation. 
  \textbf{(a)}: Quantitative evaluation shows that Any2all achieves superior perceptual quality, reflected by lower FID scores across all modalities compared to baseline methods. The Any2all results are generated using a standard DDPM sampler from a single, unconditionally trained model. 
  \textbf{(b)}: Qualitative comparison of T1 MRI and PET images generated by Any2all and M2DN (a baseline trained on mixed conditional/unconditional tasks). This figure shows that the images from Any2all exhibit finer anatomical details and fewer artifacts, visually confirming the superior quality suggested by the quantitative metrics.}
  \label{fig:unconditional}
\end{figure*}

\section{Numerical Validation}

\subsection{Dataset}
We evaluated our proposed Any2all framework using an in-house brain imaging dataset comprising T1-weighted (T1) MRI, T2-weighted FLAIR (T2FLAIR) MRI, T2-star (T2*) MRI, CT, and PET ($^{18}$F-Florbetapir) scans. Imaging data were acquired from 226 participants enrolled in an ongoing neuroimaging study on memory and aging. The study was conducted under a protocol approved by the Institutional Review Board, and all participants provided written informed consent. The median age of participants was 70 years (interquartile range [IQR]: 65--75), and 127 of the 226 participants were female. All participants underwent single time-point tri-modality imaging (PET/MR/CT). More details on data acquisition are included in the appendix. 

Raw image data for all modalities were extracted from the corresponding DICOM files. This dataset was then randomly partitioned into a training set (198 subjects), a validation set (7 subjects), and a testing set (21 subjects). For each subject, images from all modalities were co-registered to the corresponding T1-weighted MR image using a 3D rigid transformation with ANTsPy. A manual inspection was performed to ensure that all images were properly co-registered. Following registration, each MR image was normalized to a 0--1 intensity range on a per-subject basis. In contrast, the CT and PET images were normalized using parameters derived from the entire training set. These group-normalization parameters for CT and PET were then applied to the validation and testing sets. For each subject, we extracted about 100 transverse slices, containing the most relevant brain regions. The Any2all framework was validated using these 2D images.

To create a well-defined image reconstruction task, we simulated undersampled k-space measurements, $\ybm$, from the co-registered MR images, $\xbm$. This was achieved using a 1D Cartesian equispaced sampling mask, $\Mbm$, to produce an acceleration factor of 4. The imaging model is formulated as $\Abm=\Mbm\Fbm\Sbm\xbm$, where $\Fbm$ is the Fourier transform and $\Sbm$ represents the coil sensitivity maps (CSMs). We synthesized the CSMs using SigPy with a relative radius of $r=1.5$ and a coil count of $n_{\text{coils}}=8$. An additive zero-mean Gaussian noise with a standard deviation of $0.01$ was also applied to the simulated measurements.

\begin{figure*}
  \centering
  \includegraphics[width=\textwidth]{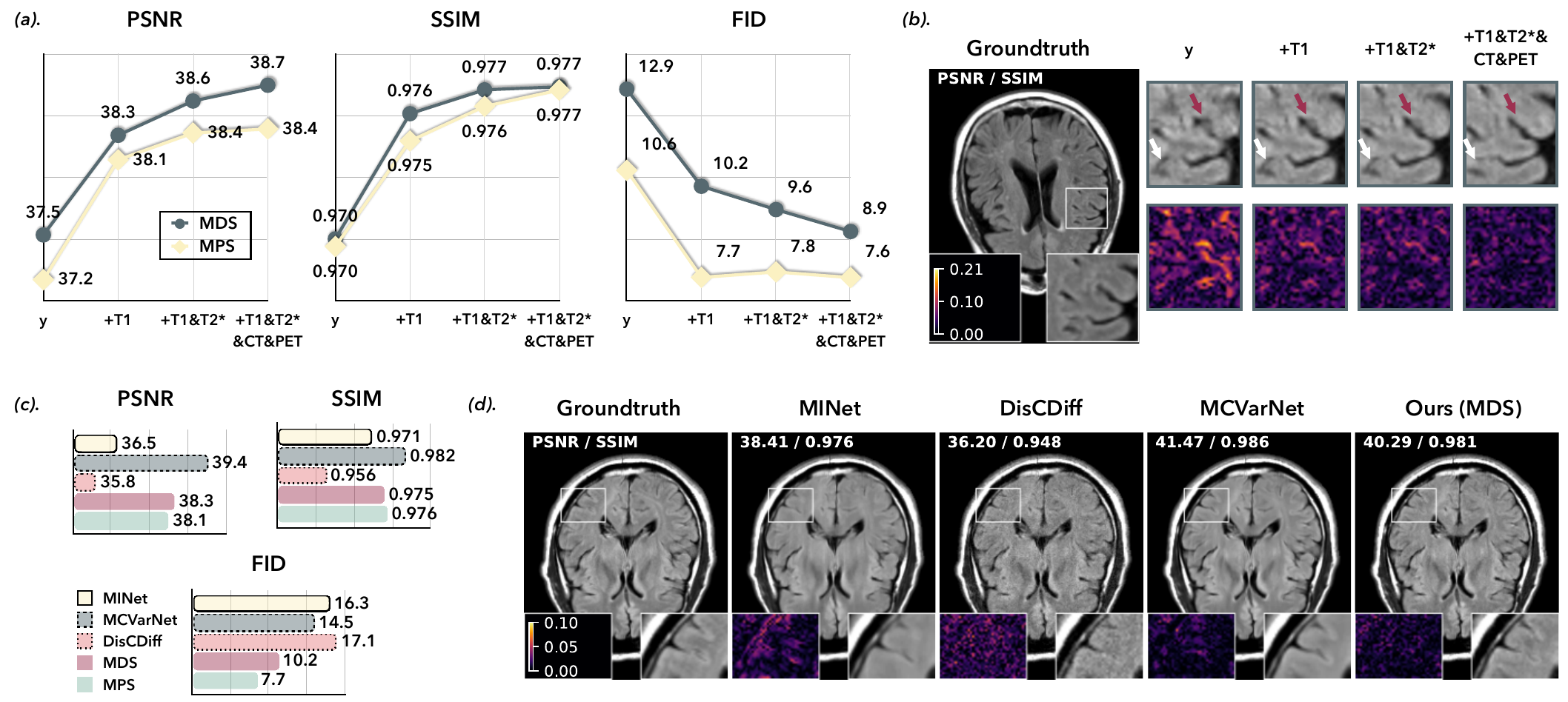}
  \caption{Qualitative and quantitative evaluation of T2FLAIR MRI reconstruction from noisy, 4x undersampled measurements, comparing Any2all with task-specific baselines. 
  \textit{\textbf{(a)}}: Quantitative comparison showing the effect of adding different auxiliary inputs. ``\emph{y}'' denotes using only the measurements. ``\emph{+T1}'', ``\emph{+T1\&T2*}'', and ``\emph{+T1\&T2*\&CT\&PET}'' indicate the addition of T1 MRI, T1 and T2* MRI, and all other modalities as auxiliary inputs, respectively.
  \textit{\textbf{(b)}}: Visual results of reconstruction using Any2all (MDS), illustrating how image quality improves with more auxiliary data. 
  \textit{\textbf{(c)}}: Quantitative comparison in the ``\emph{+T1}'' setting. Unlike Any2all, the baseline methods were trained specifically for this task. 
  \textit{\textbf{(d)}}: Visual comparison for reconstruction in the ``\emph{+T1}'' setting. 
  This figure shows that: \emph{(1)} reconstruction quality improves as more auxiliary modalities are provided; \emph{(2)} within our framework, MDS excels on distortion-based metrics while MPS achieves superior perceptual quality; and \emph{(3)} compared to baselines, Any2all achieves the best perceptual quality (lower FID) while remaining competitive in fidelity.}
  \label{fig:inverse-problem}
\end{figure*}

\subsection{Experimental Setup}
To validate the Any2all framework, we designed three distinct experiments: (1) \emph{unconditional image generation:} We evaluated the baseline quality of our trained diffusion prior by generating a complete set of multimodal images from random noise, without any conditional inputs. For this task, Any2all relies on the standard DDPM sampling algorithm~\cite{Ho.etal2020}. This experiment simulates the need for synthetic complete sets of clean, co-registered multimodal images, with potential applications in areas such as data augmentation~\cite{billot2023synthseg}, (2) \emph{multimodal image reconstruction:} We evaluated the model on the inverse problem of reconstructing a target T2FLAIR image from its undersampled k-space measurements. This experiment addresses the clinical need to accelerate T2FLAIR MRI acquisition by improving reconstruction quality with clean auxiliary images from other modalities. We systematically investigated the impact of using different auxiliary modalities as conditional inputs to guide the reconstruction, and (3) \emph{multimodal image synthesis:} We tested the model on three separate synthesis tasks: generating T2FLAIR, CT, and PET images. For T2FLAIR, this experiment simulates the synthesis of a missing MRI scan from other acquired data (MR, CT, or PET), which can be necessary for reasons such as time constraints or imaging artifacts. For CT and PET, we simulated their synthesis from various combinations of MRI data, with the potential to reduce the patient's radiation exposure. For each task, we evaluated performance when guided by different combinations of available source modalities. More details on implementation are included in the appendix.

\subsection{Comparison Methods}
We compared the performance of our Any2all framework against several state-of-the-art methods. For {multimodal image synthesis and unconditional image generation}, we compared Any2all against mmGAN~\cite{Sharma.Hamarneh2020}, mmResViT, and M2DN~\cite{Meng.etal2024a}. mmGAN is a foundational GAN architecture for multimodal MRI synthesis. mmResViT enhances mmGAN by incorporating a ResViT~\cite{Dalmaz.etal2022} generator, which uses vision transformers to better capture global image context. M2DN is a diffusion model that formulates synthesis as ``whole-modality inpainting,'' using a modality mask to handle arbitrary missing inputs. These methods are designed for ``many-to-many'' image synthesis. These methods are specific for image-to-image translation. To enable unconditional generation, we trained these models by mixing with all-noises inputs.

For {multimodal image reconstruction}, we compared Any2all against MINet~\cite{Xiang.etal2019}, MC-VarNet~\cite{Lei.etal2023}, and DisCDiff~\cite{Mao.etal2023a}. MINet is a CNN-based multi-stage network for multi-contrast MRI reconstruction. MC-VarNet is an interpretable, deep unfolding network that decomposes multi-contrast images into common and unique components to guide reconstruction. DisCDiff is a conditional diffusion model that uses disentanglement for multi-contrast MRI super-resolution. These baselines are highly task-specific. We trained these models for T2FLAIR image reconstruction using a clean T1-weighted MR image as auxiliary information.

We evaluated the performance of all methods using three standard metrics. To assess image fidelity and reconstruction accuracy, we used the \emph{peak signal-to-noise ratio (PSNR)} and the \emph{structural similarity index measure (SSIM)}. To evaluate perceptual quality and realism, we used the \emph{Fr\'echet Inception Distance (FID)}.

\begin{figure*}
  \centering
  \includegraphics[width=\textwidth]{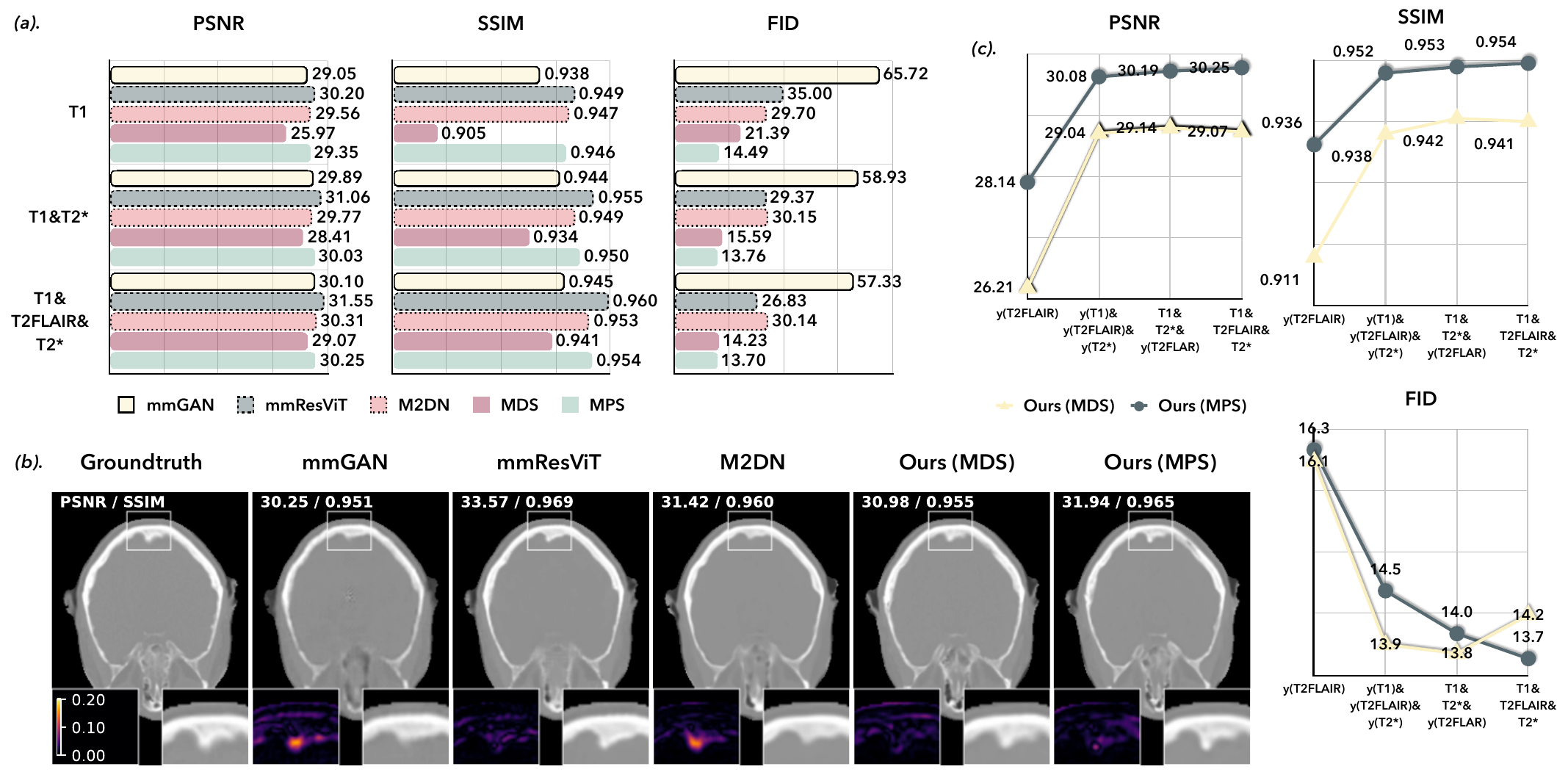}
  \caption{Qualitative and quantitative evaluation of CT image synthesis, comparing Any2all with several baseline methods designed for multimodal image synthesis.
  \textit{\textbf{(a)}}: Quantitative results for CT synthesis from different input modalities. Here, ``\emph{+T1}'' denotes synthesis from only T1 images, while ``\emph{+T1\&T2*}'' and ``\emph{+T1\&T2FLAIR\&T2*}'' indicate the addition of T2* MRI and all available MRI modalities as inputs, respectively. 
  \textit{\textbf{(b)}}: Visual results for CT synthesis given all MRI modalities as input (the ``\emph{+T1\&T2FLAIR\&T2*}'' setup).
  \textit{\textbf{(c)}}: Quantitative results for CT synthesis showcasing Any2all's unique ability to handle mixed inputs (i.e., both raw measurements and clean images). Note that the baseline methods (mmGAN, mmResViT, M2DN) only allow clean images as input. 
  This figure shows that \emph{(1)} the performance of Any2all progressively improves as more information—either from additional modalities or more complete measurements—is provided, \emph{(2)} for image synthesis, MPS excels on both distortion-based and perceptual metrics, and \emph{(3)} compared to baselines, Any2all achieves the best perceptual quality (lower FID) while remaining competitive in fidelity.
  }
  \label{fig:ct}
\end{figure*}

\subsection{Results}
\subsubsection{Unconditional Generation}
Figure~\ref{fig:unconditional} compares the performance of Any2all, mmGAN, mmResViT, and M2DN in generating realistic multimodal images from pure noise.
Figure~\ref{fig:unconditional}a presents a quantitative evaluation (FID $\downarrow$) of the generated images across all modalities. The results show that Any2all achieves lower FID scores than all baselines across all target modalities, demonstrating the superior perceptual quality of its learned generative prior.
Figure~\ref{fig:unconditional}b provides a qualitative comparison of T1 MR and PET images generated by M2DN and Any2all. The proposed Any2all model generates highly realistic samples, exhibiting fine anatomical details and negligible artifacts. Notably, Any2all produces sharp and clean images, whereas the images from M2DN suffer from blurring artifacts, particularly in the PET results.

\subsubsection{Multimodal Image Reconstruction}

Figure~\ref{fig:inverse-problem} shows a qualitative and quantitative evaluation of T2FLAIR MRI reconstruction from noisy $4\times$ undersampled measurements, comparing Any2all with several task-specific baseline methods. Here, we show results for reconstruction with various auxiliary inputs: ``\emph{y}'' denotes using only the T2FLAIR measurements; ``\emph{+T1}'', ``\emph{+T1\&T2*}'', and ``\emph{+T1\&T2*\&CT\&PET}'' indicate the addition of a T1-weighted MR image, both T1 and T2* MR images, and all other modalities as auxiliary inputs, respectively. In contrast to Any2all, the baseline methods are task-specific. We trained and tested them exclusively on the ``\emph{+T1}'' setting, which aligns with their original publications.

Figure~\ref{fig:inverse-problem}a shows a quantitative comparison demonstrating the effect of adding auxiliary modalities to Any2all. These results demonstrate two key points: \textbf{(1)} The quality of the reconstructed T2FLAIR image progressively improves as more modalities are provided; \textbf{(2)} MDS consistently outperforms MPS on distortion-based metrics, while MPS achieves superior perceptual quality. The results also highlight that the largest improvement comes from adding a single clean image from another modality (the ``\emph{+T1}'' setting).
Figure~\ref{fig:inverse-problem}b displays visual results of image reconstruction using Any2all (MDS), illustrating how image quality improves with more auxiliary data. The addition of more modalities allows Any2all to reconstruct finer details more consistently, as highlighted by the white and red arrows. These improvements can also be seen in the corresponding error maps.
Figure~\ref{fig:inverse-problem}c presents a quantitative baseline comparison for T2FLAIR reconstruction using a T1-weighted image as guidance (the ``\emph{+T1}'' setting). The results show that, within Any2all, our proposed MPS method achieves the best perceptual quality (lowest FID), while MDS remains competitive in terms of fidelity. Notably, MDS outperforms the MINet and DisCDiff baselines in both metrics and is only marginally worse than MC-VarNet on distortion-based metrics.
Figure~\ref{fig:inverse-problem}d provides a visual baseline comparison for T2FLAIR reconstruction in the ``\emph{+T1}'' setting. Our MDS method produces reconstructions with visibly sharper anatomical details and fewer aliasing artifacts compared to the baseline methods. These results also highlight that, while MC-VarNet has better PSNR and SSIM values, its results are visibly less realistic than those of Any2all (MDS) despite rendering similar anatomical details.

\subsubsection{Multimodal Image Synthesis}

Figures~\ref{fig:ct} illustrates the qualitative and quantitative evaluation of CT image synthesis, comparing Any2all with several baseline methods. In these figures, ``\emph{+T1}'' denotes synthesis from only T1 images, while ``\emph{+T1\&T2*}'' and ``\emph{+T1\&T2FLAIR\&T2*}'' indicate the addition of T2* MRI and all available MRI modalities as inputs, respectively. All baseline methods are designed for multimodal image synthesis; unlike Any2all, they do not exhibit the ability for multimodal reconstruction or accept mixed inputs. We also tested Any2all with mixed inputs (i.e., both raw measurements and clean images), using \emph{y($\cdot$)} to represent the 4x undersampled and noisy measurements of an MRI modality. For example, \emph{``y(T1)''} denotes the measurements of the T1 MRI data. More results on PET and T2FLAIR image synthesis are included in the appendix.

Figure~\ref{fig:ct}a presents quantitative results for CT synthesis from different input modalities. The results demonstrate that Any2all, using either MDS or MPS, achieves the best perceptual quality and competitive performance on distortion-based metrics across these setups. Within Any2all, MPS consistently outperforms MDS across all settings and metrics. Furthermore, the synthesis quality for both MPS and MDS improves as more input modalities are provided.
Figure~\ref{fig:ct}b shows visual results for CT synthesis given all MRI modalities as input (the ``\emph{+T1\&T2FLAIR\&T2*}'' setup). The images synthesized by Any2all, using both MPS and MDS, are visually competitive with those produced by the baseline methods and display fine structural details.
Figure~\ref{fig:ct}c presents quantitative results for CT synthesis that showcase Any2all's unique ability to handle mixed inputs. Note that the baseline methods (mmGAN, mmResViT, M2DN) only allow clean images as input. These results show that the performance of Any2all (MPS) progressively improves as more information—either from additional modalities or more complete measurements—is provided. It is also worth noting that we empirically observed the improvement between ``\emph{y(T2FLAIR)}'' and ``\emph{y(T1)}+\emph{y(T2FLAIR)}+\emph{y(T2*)}'' is more significant than that between ``\emph{y(T1)}+\emph{y(T2FLAIR)}+\emph{y(T2*)}'' and ``\emph{T1}+\emph{y(T2FLAIR)}+\emph{T2*}''. This indicates that increasing the number of available modalities is more impactful than improving the information quality of specific modalities (i.e., from noisy measurements to clean images).

\section{Conclusion and Discussion}

In this paper, we introduced \emph{Any2all}, a unified generative framework designed to overcome the complexity and inflexibility of conventional multimodal medical imaging. Our core contribution is a novel formulation that reframes the disparate tasks of image reconstruction and synthesis into a single, coherent \emph{virtual inpainting problem}. This unified viewpoint allows a single, unconditionally trained diffusion model to address a comprehensive range of imaging scenarios. To realize this, we proposed two inference algorithms, namely MPS and MDS to guide the generation process using any available combination of input data.

Our extensive validation on a clinical MRI, CT, and PET dataset confirms the power and versatility of the Any2all framework. The experiments demonstrated that our model learns a high-quality generative prior, capable of producing realistic, high-fidelity images from pure noise. In task-specific evaluations, Any2all consistently achieves competitive distortion-based performance and state-of-the-art perceptual performance. For multimodal reconstruction, it outperforms specialized baselines, exhibiting a compelling trade-off between the superior perceptual quality of MPS and the high fidelity of MDS. For multimodal synthesis, the framework's flexibility was proven across multiple target modalities (T2FLAIR, CT, and PET), where it again produced images with perceptual quality that was significantly better than competing methods. The results robustly show that the quality of the generated images consistently improves as more auxiliary information is provided, validating the model's ability to effectively fuse multimodal data.

The primary implication of this work is the potential to dramatically simplify clinical and research workflows by replacing an entire suite of task-specific models with a single, flexible, and powerful framework. This can reduce the burden of model development, training, and maintenance. However, we also acknowledge limitations that suggest avenues for future work. A trade-off exists between our sampling methods, where MPS is preferable for perceptual realism and MDS for quantitative fidelity in multimodal image reconstruction. Furthermore, the iterative nature of diffusion models, while yielding high-quality results, is more computationally intensive at inference time than one-shot GAN-based methods.

Future research could focus on integrating faster sampling techniques to improve clinical feasibility. Extending and validating the Any2all framework on other anatomical regions, additional imaging modalities, and other inverse problems like deblurring would be a valuable next step. Ultimately, this work establishes the virtual inpainting approach as a potent and generalizable paradigm for multimodal imaging, holding great promise for enhancing the efficiency and capability of medical image analysis.

\section*{Acknowledgments}

Research presented in this article was supported by the NSF CAREER award CCF-2043134. This work was also supported by the NSF Award CCF-2504613.

\section*{Impact Statement}

This paper presents work whose goal is to advance the field of Machine
Learning. There are many potential societal consequences of our work, none
which we feel must be specifically highlighted here.

\newpage
\appendix
\onecolumn
\section{Appendix}

\subsection{Data acquisition}

PET and MR images were acquired using an integrated Biograph mMR PET/MRI system (Siemens AG, Erlangen, Germany). $^{18}$F-Florbetapir (Amyvid, Eli Lilly, Indianapolis, IN) PET data were acquired with a median injection dose of 377.4 MBq (IQR: 362.6--384.8). The PET data were reconstructed into four distinct temporal phases; however, only images from the first phase were used in subsequent experiments. 
MR acquisition parameters were as follows. T1-weighted images were acquired using a 3D MPRAGE sequence with: TE$/$TR $= 2.95/2300$ ms, TI $= 900$ ms, number of partitions $= 176$, matrix size $=240\times 256 \times 176$, voxel size $= 1.1\times1.1 \times 1.2$ mm$^3$, and acquisition time $= 5$ min $12$ s. T2-weighted FLAIR images were acquired using a 2D sequence with: TE$/$TR $= 91/9000$ ms, TI $= 2500$ ms, matrix size $=256\times 256 \times 35$, voxel size $=0.9\times 0.9\times 5$ mm$^3$, and acquisition time $= 4$ min $05$ s. T2-star images were acquired using a 2D sequence with: TE$/$TR $=20/650$ ms, matrix size $=256\times 256 \times 44$, voxel size $= 0.8 \times 0.8 \times 4.0$ mm$^3$, and acquisition time = $4$ min $11$ s. During the study period, the Siemens mMR system was upgraded from VB20P to VE11P, which included software and scanner computer updates. A total of 85 PET/MR scans were performed on the upgraded Syngo VE11P system, while the remaining scans were acquired using the Syngo VB20P.

CT images were acquired using a Biograph 40 PET/CT system (Siemens AG, Erlangen, Germany). The median interval between CT and PET/MRI acquisitions was 0 days (IQR: -6 to 22). CT images were acquired at $120$ kVp with a matrix size of $512 \times 512 \times 74$ and a voxel size of either $0.59 \times 0.59 \times 3.0$ mm$^3$ or $0.59 \times 0.59 \times 2.0$ mm$^3$.

\subsection{Implementation}

We implemented Any2all using the publicly available \textsf{guided-diffusion} codebase~\cite{dhariwal2021diffusion}. All experiments were performed on a machine equipped with an AMD EPYC 7443P Processor and four NVIDIA A6000 GPUs. For training, we set the number of diffusion timesteps to 1000 and used a cosine noise schedule as proposed in~\cite{dhariwal2021diffusion}. The model was trained for 1.5 million iterations with a batch size of 24, using the Adam optimizer with a learning rate of $1 \times 10^{-4}$. The total training time was approximately two weeks. For sampling, we used 1000 steps for MPS and 150 steps for MDS.
Table~\ref{tb:implementation} summarizes the publicly available software used in this study. All other methods were implemented by our team.

\subsection{Additional Results}
\begin{figure*}
  \centering
  \includegraphics[width=\textwidth]{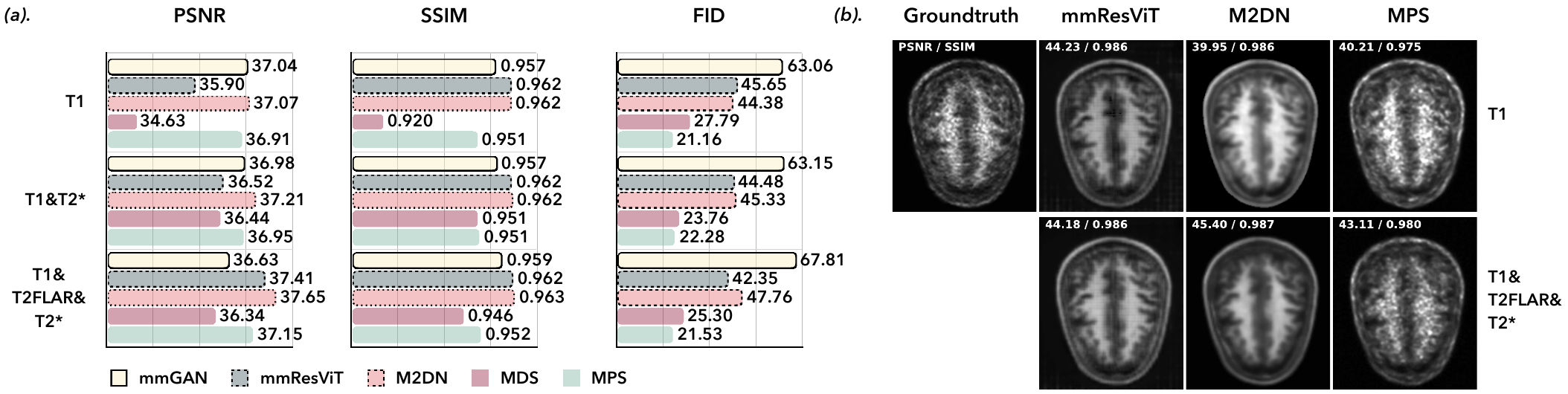}
  \caption{Qualitative and quantitative evaluation of PET image synthesis, comparing Any2all with several baseline methods designed for multimodal image synthesis. 
  \textit{\textbf{(a)}}: Quantitative results for PET image synthesis. Here, ``\emph{+T1}'' denotes synthesis from only T1 images, while ``\emph{+T1\&T2*}'' and ``\emph{+T1\&T2FLAIR\&T2*}'' indicate the addition of T2* MRI and all available MRI modalities as inputs, respectively. 
  \textit{\textbf{(b)}}: Visual results from Any2all (MPS), M2DN, and mmResViT for PET synthesis, given either T1 MR images or all MRI modalities as input. This figure shows that \emph{(1)} the performance of Any2all progressively improves as more information is provided, \emph{(2)} for image synthesis, MPS excels on both distortion-based and perceptual metrics, and \emph{(3)} compared to baselines, Any2all achieves the best perceptual quality (lower FID) while remaining competitive in fidelity. Note how MPS generates images with a perceptual quality that consistently matches the ground truth, whereas the baseline methods produce oversmoothed results that lack fine details.}
  \label{fig:pet}
\end{figure*}

\begin{figure*}
  \centering
  \includegraphics[width=\textwidth]{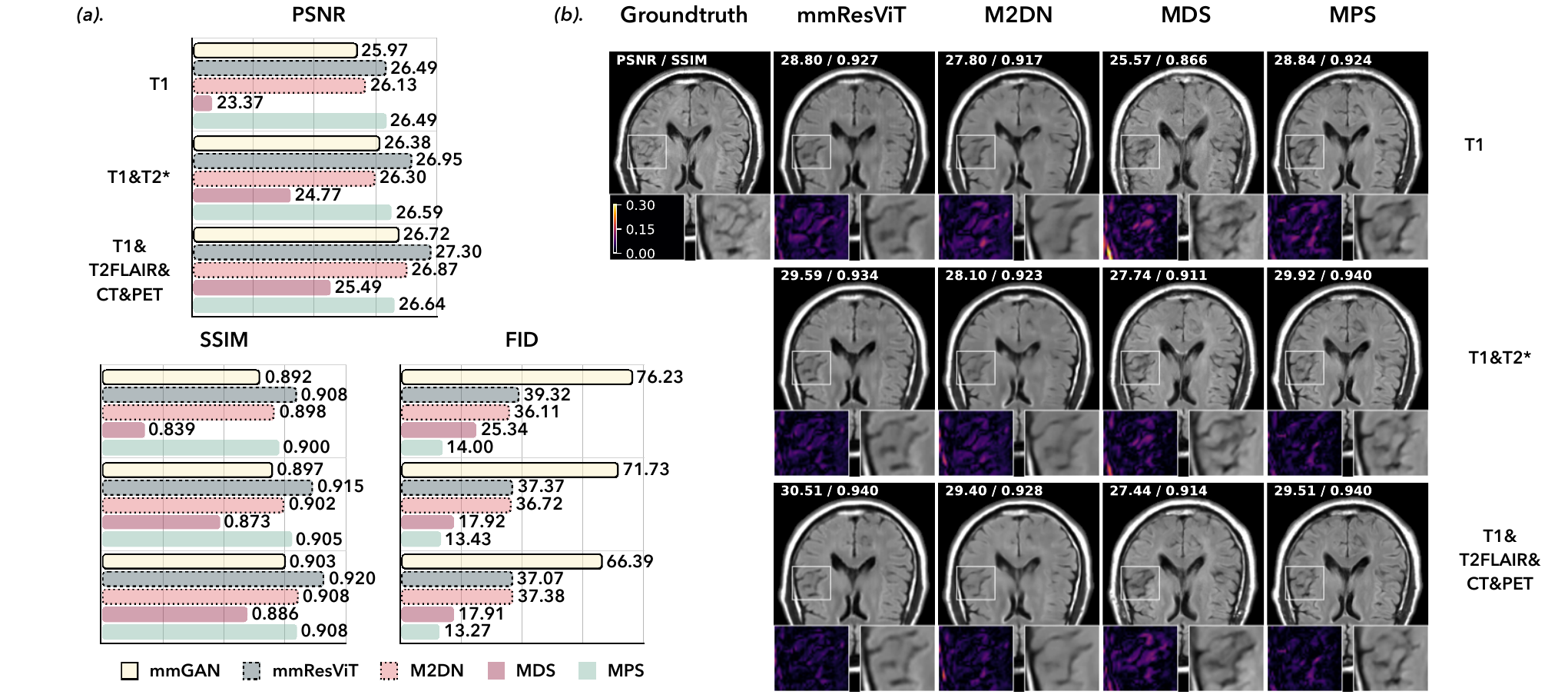}
  \caption{Qualitative and quantitative evaluation of T2FLAIR MR image synthesis, comparing Any2all with several baseline methods designed for multimodal image synthesis.
  \textit{\textbf{(a)}}: Quantitative results for T2FLAIR synthesis. Here, ``\emph{+T1}'' denotes synthesis from only T1 images, while ``\emph{+T1\&T2*}'' and ``\emph{+T1\&T2FLAIR\&CT\&PET}'' indicate the addition of T2* MRI and all other modalities as inputs, respectively. 
  \textit{\textbf{(b)}}: Visual results for T2FLAIR synthesis. This figure shows that \emph{(1)} the performance of Any2all progressively improves as more information is provided, \emph{(2)} for image synthesis, MPS excels on both distortion-based and perceptual metrics, and \emph{(3)} compared to baselines, Any2all achieves the best perceptual quality while remaining competitive in fidelity.}
  \label{fig:t2flair_synthesis}
\end{figure*}

Figures~\ref{fig:pet} and~\ref{fig:t2flair_synthesis} illustrate the qualitative and quantitative evaluation of PET and T2FLAIR image synthesis, respectively, comparing Any2all with several baseline methods.

Figure~\ref{fig:pet}a presents quantitative results for PET synthesis. Our MPS method achieves a competitive overall performance, excelling in both fidelity (PSNR/SSIM) and perceptual quality (lowest FID) across most input scenarios. The results also show that the quality of images synthesized by both MPS and MDS quantitatively improves as more modalities are provided as input.
Figure~\ref{fig:pet}b shows visual results from Any2all (MPS), M2DN, and mmResViT for PET synthesis, given either T1-weighted MR images or all MRI modalities as input. The images generated by MPS have a perceptual quality that consistently matches the ground truth, whereas the baseline methods produce oversmoothed results that lack fine detail.

Figure~\ref{fig:t2flair_synthesis}a shows quantitative results for T2FLAIR synthesis. Our MPS method achieves a competitive performance, showing strong results in both fidelity (PSNR/SSIM) and perceptual quality (lowest FID) across most input scenarios. The results also confirm that the quality of images synthesized by both MPS and MDS quantitatively improves as more modalities are provided. Notably, MPS achieves better performance than the M2DN diffusion baseline in both the ``\emph{+T1}'' and ``\emph{+T1\&T2*}'' settings.
Figure~\ref{fig:t2flair_synthesis}b presents visual results for T2FLAIR synthesis. The images generated by our methods, particularly MPS, exhibit sharper anatomical structures and fewer artifacts compared to the baseline methods. As more input modalities are included, the synthesized T2FLAIR images become visibly more detailed and structurally coherent. For example, MPS produces sharp imaging details in the ``\emph{+T1}'' setting, and these details become richer when more modalities are included as input.

\begin{table*}
  \small
  \centering
  \caption{Publicly available implementations used in this study.}
  \label{tb:implementation}
  \begin{tabular}{ll}
    \toprule
  Method & Public implementation: $\mathsf{https://github.com/}$ \\
  \midrule
      ANTsPy   & ANTsX/ANTsPy \\
      SigPy & mikgroup/sigpy \\
      MINet~\cite{Xiang.etal2019} & chunmeifeng/MINet \\
      MC-VarNet~\cite{Lei.etal2023} & lpcccc-cv/MC-VarNet \\
      DisCDiff~\cite{Mao.etal2023a} & Yebulabula/DisC-Diff \\
      mmGAN~\cite{Sharma.Hamarneh2020} & trane293/mm-gan \\
      ResViT~\cite{Dalmaz.etal2022} & icon-lab/ResViT \\
      guided-diffusion~\cite{dhariwal2021diffusion} & openai/guided-diffusion \\
    \bottomrule    
  \end{tabular}
\end{table*}

\end{document}